\documentclass{appolb}

\usepackage{graphicx}
\usepackage{cite}
\usepackage{amsmath}

\def\slashchar#1{\setbox0=\hbox{$#1$}
   \dimen0=\wd0 \setbox1=\hbox{/} \dimen1=\wd1
   \ifdim\dimen0>\dimen1 \rlap{\hbox to \dimen0{\hfil/\hfil}} #1
   \else  \rlap{\hbox to \dimen1{\hfil$#1$\hfil}} / \fi}

\begin{document}
\title{Chiral waves in quark matter% - a remake%
\thanks{Presented at HIC for FAIR Workshop, XXVIII Max Born Symposium
{\em Three days on Quarkyonic Island}, Wroc\l{}aw, 19-21 May 2011}
\thanks{Supported in part by the Polish Ministry of Science and Higher Education, grants N~N202~263438 and N~N202~249235}
}
\author{Wojciech Broniowski
\address{Institute of Physics, Jan Kochanowski University, PL-25406~Kielce, Poland, and\\ 
Institute of Nuclear Physics PAN, PL-31342~Cracow, Poland}
%\and
%the Name(s) of other Author(s)
%\address{and their affiliation}
}

\date{18 October 2011}

\maketitle

\begin{abstract}
A mini-review of non-uniform phases in quark matter is presented, with 
particular attention to the pion condensation, also known as chiral density 
waves or chiral spirals. The phase diagram of strongly-interacting matter 
may involve such a phase, placed on the {\em quarkyonic island} between the baryonic phase and the chirally-restored 
quark-gluon plasma. 
\end{abstract}

  
\section{Introduction}
The goal of this talk is to present a short ``historical'' review of 
various aspects of non-uniform phases in quark matter, which at present are 
being actively investigated in a modern framework of the {\em quarkyonic} medium. 

\begin{figure}[tb]
\centerline{\includegraphics[width=.73\textwidth]{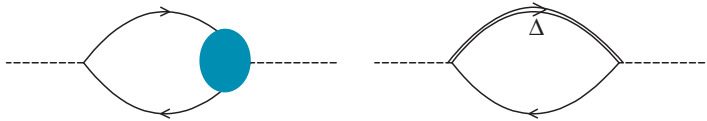}} 
\caption{The particle-hole and $\Delta$ excitations in the pion self-energy in nuclear medium. \label{fig:phD}}
\end{figure}

As is well known, the pion emerges as a pseudo-Goldstone boson of the spontaneously broken chiral symmetry. The attractive $P$-wave interaction with 
fermions (nucleons, quarks) opens the possibility of the pion condensation in nuclear matter. The phenomenon was first 
explored in the work by Migdal~\cite{Migdal:1971cu,Migdal:1973zm}, where the particle-hole and $\Delta$ resonance effects of Fig.~\ref{fig:phD} were
included in the pion self-energy non-relativistic calculation. Numerous works followed the issue of the pion condensation
in nuclear matter (for reviews and references 
see~\cite{Baym:1974,Brown:1976,Baym:1979,Kunihiro:1993pz}). A recent status is reported in~\cite{Akmal:1997ft}, where the appearance of the pion 
condensation is expected at $\sim 1.5-2$ nuclear saturation density for the symmetric and neutron matter, however, the results
depend in a quite sensitive way on the parameters of the $NN$ interactions. 

We mention above the ``ancient'' works, as the questions concerning non-uniform phases in quark matter are very much related to 
the old issue of the pion condensation. 
As a matter of fact, a similar phenomenon, linked
to particle-hole superconductivity in strong spin-exchange fields has been known in condensed matter
for a long time~\cite{Larkin:1964,Fulde:1964}, leading to Alternating Layer Spin structures (see, e.g. \cite{Tatsumi:1980aa} and references therein). 

\section{Pion condensation in a relativistic approach \label{sec:relativ}}

\begin{figure}[tb]
\centerline{\includegraphics[width=.43\textwidth]{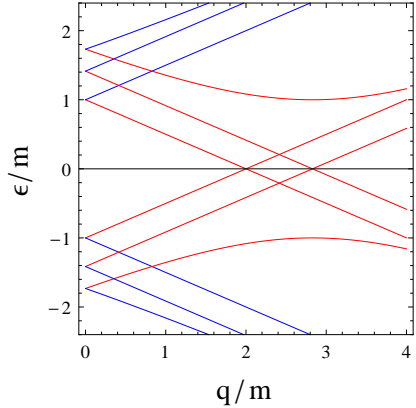}} 
\caption{The Dirac spectrum  (\ref{eq:sp}) in the periodic chiral field ansatz (\ref{eq:ans}), plotted for 
sample values of the quark momentum $\vec{k}$. We note the appearance of two branches for the positive-energy 
spectrum. The lower branch is for  $u \downarrow$ and $d \uparrow $ states, while the upper one is 
for $u \uparrow$ and $d \downarrow $ states. \label{fig:eps}}
\end{figure}

In 1979 Dautry and Nyman~\cite{Dautry:1979bk} found an analytic solution of the relativistic problem of the pion 
condensation, using the $\sigma$-model Lagrangian
\begin{eqnarray}
L = \bar \psi \left ( i \slashchar \!\!\!\! \partial -g(\sigma+ i \tau \cdot {\boldsymbol \pi}) \right ) \psi + \dots \label{eq:L}
\end{eqnarray}
where $\psi$ was the Dirac nucleon field (we change it to quarks), ellipses denote the meson 
kinetic and interaction terms, $g$ is a coupling constant, 
finally the sigma and the pion mean fields assume the periodic ansatz 
\begin{eqnarray}
\sigma = f \cos[q(z-z_0)], \;\; \pi^0=f \sin[q(z-z_0)], \;\; \pi^\pm=0, \label{eq:ans}
\end{eqnarray}
with $f=93$~MeV being the pion decay constant. The wave vector $\vec{q}$, here pointing in the $z$-direction, is a parameter, and 
$z_0$ is arbitrary.
The Dirac spectrum of (\ref{eq:L}) has the form~\cite{Dautry:1979bk} 
\begin{eqnarray}
\epsilon(k) = \pm \sqrt{M^2+k^2+q^2/4 \pm \sqrt{M^2 q^2 +q\cdot k^2}}, \label{eq:sp}
\end{eqnarray}
where $\vec{k}$ is the quark momentum, 
the outer $\pm$ sign denotes the positive and negative energy solutions, while the $\pm$ sign inside the square root 
defines the branch. The spectrum (\ref{eq:sp}) for various fixed values 
of $\vec{k}$, is visualized in Fig.~\ref{fig:eps}. We note that with the increasing value of $q$ the branches 
split, with one increasing, and the other decreasing in energy. Thus, Fermi-sea energy may be lowered when the lower branch is 
occupied. 

Similarly, the ansatz for the charged pion condensation has the form~\cite{Dautry:1979bk}
\begin{eqnarray}
\pi_1 = f \cos[q (z-z_0)], \;\; \pi_2=f \sin[q (z-z_0)], \;\; \pi_3=\sigma=0, \label{eq:ansc}
\end{eqnarray}
which is obtained with the chiral rotation $\exp \left ( i {\frac{\pi}{4}} \gamma_5 \tau_1 \right )$ from (\ref{eq:ans}), 
therefore yields the same spectrum and thermodynamic features, however, offers no interesting magnetic properties (see Sec.~\ref{sec:mag}).
We note here some resemblance to the issue of {\em disoriented chiral condensates} (for a review and reference see, e.g., \cite{Mohanty:2005mv}):
forms (\ref{eq:ans},\ref{eq:ansc}) may be viewed as ``periodic disoriented chiral condensates''.

To understand the dynamical mechanism behind the formation of the pion 
condensate in more general terms, we may carry out the chiral rotation
$\psi = e^{-i\gamma_5 /(2f) {\boldsymbol \tau} \cdot {\boldsymbol \phi}} q$ on Lagrangian~(\ref{eq:L}), which yields
\begin{eqnarray}
{\cal L}= \bar q \left (  i \slashchar \!\!\!\! \partial - 
\frac{1}{2f}\gamma_5 \gamma_\mu (\partial^\mu {\boldsymbol \phi}) \cdot {\boldsymbol \tau} - M  \right ) q + \dots,
\end{eqnarray}
with $M=gf$ denoting the constituent quark mass due to the spontaneous breaking of the chiral symmetry.
Attraction occurs for fermions (quarks) with appropriately correlated spin $\Sigma^i=\gamma_5 \gamma^0 \gamma^i$ and flavor
whenever the pion field is nonuniform. For the ansatz (\ref{eq:ans}) we obtain attraction for $|u\downarrow \rangle$ and  
$|d\uparrow \rangle$ states (the decreasing positive-energy branch in Fig.~\ref{fig:eps}), while repulsion is found for 
$|u\uparrow \rangle$ and  $|d\downarrow \rangle$ (the rising positive-energy branch). This is reminiscent of the well-known hedgehog 
form found in chiral bags~\cite{Chodos:1975ix} and chiral solitons~\cite{Birse:1983gm,Kahana:1984dx}, where 
the valence quarks occupy the correlated spin-flavor state $q_h=1/\sqrt{2} (|u\downarrow \rangle - |d\uparrow \rangle)$. 

The pion condensation is driven by the Fermi sea, favoring increased values of $q$. At the same time, the meson kinetic terms 
tend to suppress $q$ (for the $\sigma$-model they give 
the contribution $\frac{1}{2} f^2 q^2$ to the energy density). Hence the occurrence of the pion-condensed
state (state with $q \neq 0$) is a rather subtle dynamical issue, here involving the Fermi energy and 
the value of the coupling constant $g$ (or, equivalently, the quark mass $M=g f$).  

\section{Pion condensation in quark matter \label{sec:pi_qm}}

\begin{figure}[tb]
\centerline{\includegraphics[width=.5\textwidth]{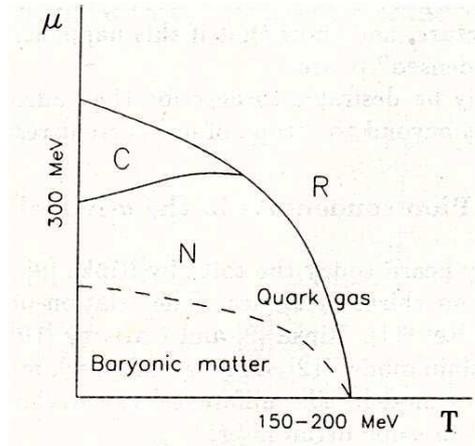}} 
\caption{The phase diagram with the quark pion-condensed phase (C), the normal quark phase with broken
chiral symmetry (N), the restored phase (R), and the baryonic phase 
(reprinted from \cite{Broniowski:1990dy}). \label{fig:phase}}
\end{figure}

In an early series of papers Kutschera, Kotlorz, and the present 
author~\cite{Kutschera:1989yz,Broniowski:1990dy,Kutschera:1990xk} explored various aspects 
of the pion condensation in quark matter. We would like to recall 
a figure and a quote from one of these papers~\cite{Broniowski:1990dy}. Figure~\ref{fig:phase} shows the result of the 
phase-diagram calculation in the $\sigma$-model. We note the appearance of the new phase (C), where 
(neutral) pion condensation appears - a spatially nonuniform phase with fields (\ref{eq:ans}).  

We have also provided a percolation argument which may support the validity of using the quark gas 
at larger baryochemical potentials $\mu$~\cite{Broniowski:1990dy}: 

\medskip

\noindent
{\em We can see that there is room 
for the new phase only if (at a given temperature) ``declustering'' occurs at a lower density than the chiral
restoration. This situation may be viewed as follows: at low densities we have isolated hadrons. As the density 
is increased, the ``bags'' start to overlap, and the quarks can percolate. This is a geometrical effect, and it 
is hard to imagine why it should occur at the same density as the chiral restoration, which is a dynamical effect.
Thus, it is possible that there exists a quark gas phase with broken chiral symmetry \dots if this happens, then
the system develops a ``pion condensed'' phase.} 

\medskip

\noindent
Essentially, this was the same simple argument as brought up recently in~\cite{Castorina:2010gy}. Another early 
work on nonuniform phase on chiral quark models, ``Standing wave ground state \dots '' was reported in
\cite{Deryagin:1992rw}.
Large-$N_c$ arguments for finite density QCD were subsequently made in~\cite{Shuster:1999tn}, while 
the possibility of the particle-hole pairing (the Overhauser effect) was studied in~\cite{Park:1999bz,Rapp:2000zd}.

The pion condensation effect may occur in the Nambu-Jona--Lasinio model as well, but it becomes quite 
sensitive to the details of the model (regularization, parameters)~\cite{Broniowski:1990gb}.
Essentially, compared to the $\sigma$-model, the NJL model replaces the meson kinetic term $\frac{1}{2} q^2 f^2$ with 
a more general function $K(q)=\frac{1}{2} q^2 f^2 + {\cal O}(q^4)$. Since $q$ is not small, the formally subleading terms are
relevant. 

A more detailed study of the pion condensation in the NJL model was first made 
by Sadzikowski and the author in~\cite{Sadzikowski:2000ap}.
Several other calculations followed, confirming the possibility of ``chiral spiral''~\cite{Schon:2000he} a.k.a.
``dual chiral density wave''~\cite{Nakano:2004cd} in the Gross-Neveu and NJL models. 
Other NJL analyses were reported in~\cite{Casalbuoni:2005zp,Partyka:2008sv,Basar:2009fg,Ebert:2011rg}, also 
in the presence of the superconducting phase~\cite{Sadzikowski:2002iy,Sadzikowski:2006jq,Inui:2007zc,Partyka:2010gk}.
Bringoltz~\cite{Bringoltz:2006pz} pointed out the possibility of the effect in strongly-coupled QCD on the lattice.
More recently, in the framework of quarkyonic matter~\cite{McLerran:2007qj} inspired with the Polyakov-loop arguments~\cite{Fukushima:2003fw,Megias:2004kc,Megias:2004hj,Fukushima:2008wg}, the
chiral spirals and their implication for the phase diagram have been 
actively investigated~\cite{Kojo:2009ha,Partyka:2010em,Carignano:2010ac,Kojo:2011cn}.

\section{Magnetization \label{sec:mag}}

A very intriguing aspect of the neutral pion-condensed phase is its magnetization property. 
For the neutral pion condensation ansatz the interaction term in the Dirac Hamiltonian can be written as 
\begin{eqnarray}
 -\frac{1}{2} \vec{\Sigma} \cdot \vec{q} \tau_3, \;\;\; \Sigma^i =\gamma_5 \gamma^0 \gamma^i = 
\left ( \begin{array}{cc} \sigma^i & 0 \\ 0 & \sigma^i \end{array} \right ),
\end{eqnarray}
i.e., attraction occurs for $|u \downarrow \rangle$ and $|d \uparrow \rangle$, 
and repulsion for $|u \uparrow \rangle$ and $|d \downarrow \rangle$ states.
By definition, the magnetization is equal to ${\cal M}=g(\mu_u s_u +\mu_d s_d)$, where $g=2$ and $\mu_u=-2 \mu_d$, 
hence
\begin{eqnarray}
{\cal M}=2 \left [ \mu_u \frac{1}{2} (n_{u \uparrow} - n_{u \downarrow}) + \mu_d \frac{1}{2} (n_{d \uparrow} - n_{d \downarrow}) \right ]
\end{eqnarray}
In flavor-symmetric matter $n_{u \uparrow}=n_{d \downarrow}$, $n_{d \uparrow}=n_{u \downarrow}$, therefore one finds
\mbox{${\cal M} =3\mu_d (n_{d \uparrow}-n_{d \downarrow})$}.
Because the $u \downarrow / d \uparrow $ branch is filled more than the $u \uparrow / d \downarrow $ branch
(cf. Fig.~\ref{fig:eps}) , $n_{d \uparrow}> n_{d \downarrow}$, and permanent magnetization occurs.
At the mean-field level one can show from stationarity that~\cite{Baym:1974}
\begin{eqnarray}
s_{u/d} = \mp \frac{1}{2} f^2 q. 
\end{eqnarray}
Hence, appearance of the chiral wave in the system is equivalent to net magnetization.
Kotlorz and Kutschera~\cite{Kotlorz:1994sk} applied the above effect to 
obtain large magnetic fields of stars with a pion-condensed quark core, up to 
$\sim 10^{15}$~G. 

Conversely, an external magnetic field induces chiral waves. Recently, interesting studies of this effect
in dense quark matter were made in~\cite{Takahashi:2007qu,Frolov:2010wn,Basar:2010zd,Tatsumi:2011tu,Frolov:2011zz,Rabhi:2011mj}, and also
for the color superconducting phases in~\cite{Ferrer:2007iw}.

\section{One-loop physics and validity of the mean-field}

A quite remarkable feature of the spectrum (\ref{eq:sp}) is that the expressions for baryon and energy 
densities (at $T=0$) are also analytic~\cite{Kutschera:1989yz,Broniowski:1990dy,Kutschera:1990xk},
which allows for an exact study of the one-quark-loop~\cite{Broniowski:1989tw} and also the one-meson-loop~\cite{Broniowski:1989fz}
contributions to the effective action for the case of a nonuniform periodic background of Eq.~(\ref{eq:ans}) (see also~\cite{Kato:1992iw}). 
This made possible a study of the convergence radius in the gradient and the heat-kernel expansions, providing interesting formal 
features. In particular, for the fermion loop the gradient expansion works for $q < 2m$, while for the
expansion in $\slashchar \!\!\!\! \partial U$, with $U$ denoting the chiral field, 
the effective expansion variable is $q^2/(4m^2 + q^2)$. For the mean-field approach to be valid, the wave vector $q$, corresponding to the 
size of the gradient of the field, should not be too large. 
%However, the one-loop corrections may be incorporated, suggesting that $q < 2m$ is the validity domain of the 
%mean-field treatment. For $m \sim 300$~MeV this provides a limit $q < 600$~MeV, or the corresponding 
%characteristic size $L = 1/q > 0.3$~fm.

\section{What is the lowest state?}

The fundamental question of {\em what the phase structure of strongly interacting matter is}  
clearly poses difficulties, as it involves dynamics of strong interactions in a multi-scale system 
of dense medium. 
Certainly, finding lower-energy solutions than a uniform state shows that the uniform state is 
unstable. Another issue, however, is if the simple ansatz of Eq.~(\ref{eq:ans}), or another 
possible solution we may find, is indeed the lowest state of the system in certain thermodynamic
conditions. 
It should be noted here that on general grounds one-dimensional structures are unstable with 
respect to thermodynamic fluctuations ~\cite{Baym:1982ca}. Hence the chiral wave, 
with a one-dimensional order parameter $q$, is 
thermodynamically unstable and there must exist yet lower energy states.

Another class of solutions with energy even lower than the chiral wave are the periodic systems of 
domain-wall solitons, investigated by Nickel and Bubbala~\cite{Nickel:2008ng,Nickel:2009ke,Nickel:2009wj,Buballa:2009ct}
by implanting solutions from the Gross-Neveu model~\cite{Schnetz:2004vr,Schnetz:2005ih} to the 3+1 dimensional case. Unlike the chiral-wave
case, where the baryon density is uniform, the periodic domain-wall solutions have a nonuniform baryon 
density, thus baryon clusterization may be studied in this case. At low densities one finds isolated 
baryon slabs, while at increasing density they start to overlap and finally dissolve. The solution 
has an interesting phase diagram, with the critical end-point coinciding with the Lifshitz point~\cite{Nickel:2009wj}.

We also comment that the inclusion of the current quark mass, or a pion mass, 
in chiral waves is not trivial if self-consistency of the solution is to be preserved.
For instance, in the $\sigma$ model one of the Euler-Lagrange equations has the form
\begin{eqnarray}
 \Box \sigma = g  \langle \bar \psi \psi \rangle + 2V'(\sigma^2+\pi^2) \sigma + {f m_\pi^2},
\end{eqnarray}
with $V'$ denoting the derivative of the chirally-symmetric Mexican Hat potential with respect to its argument. The constant 
term coming from the explicit breaking of the chiral symmetry, $f m_\pi^2$, 
breaks the ansatz (\ref{eq:ans}), hence for nonzero $m_\pi$ the Dautry-Nyman solution is not consistent.
A study by Maedan~\cite{Maedan:2009yi}, where an expansion in $m_\pi$ is applied, 
shows that the chiral waves survive the nonzero-$m_\pi$ corrections.

At the same time we know that the inclusion of a nonzero pion mass has significant effects on the phase diagram. In particular, 
the location of the critical point is sensitive to $m_\pi$. This fact has been pointed out 
for the uniform solution~\cite{Scavenius:2000qd} and for the periodic domain-wall solution.

\section{Conclusion}

We have a mounting evidence that quark matter likes to form spatially non-uniform structures. Apart for the discussed 
solutions we also have Skyrmion crystals~\cite{Klebanov:1985qi,Goldhaber:1987pb,Castillejo:1989hq,Kugler:1988mu}, not 
discussed here. Also, crystalline structures may form in the high-density color superconducting 
phases~\cite{Alford:2000ze}. On the other hand, at low densities we know that baryons are the proper degrees 
of freedom. With all this in mind it is natural to expect that the transition from the baryonic phase 
to the (uniform) quark-gluon plasma may occur via non-uniform phases. This is somewhat reminiscent of the formation 
of nuclear pasta~\cite{Pethick:1995di} at baryon densities below nuclear saturation density.

Nonuniformity of the medium poses a challenge, as analytic methods are of limited application; known 
example discussed above concern certain periodic systems. Also, important details (existence of new phases, 
nature of phase transitions) depend sensitively on the model parameters whose values are difficult to assess   
in the medium. Our knowledge, or rather the spectrum of possibilities for the structure of dense medium
comes largely from effective quark models (for a review see, e.g.,~\cite{Schaefer:2007pw}), with parameters set 
from the properties in the vacuum sector. Thus, the uncertainties are large. 

The pion attraction plays a dominant dynamical role in the physics of strong interactions and, as 
stressed in Sect.~\ref{sec:relativ}, it drives the system towards chiral waves, 
making the uniform phase unstable in the domain indicated in Fig.~\ref{fig:phase}. 
We expect strong spin-isospin correlations in this phase, leading to interesting magnetic effects. 

We have also argued throughout this talk that the following terms are ``historically'' linked: 

\medskip

\noindent
{\em pion condensation $\sim$ particle-hole pairing $\sim$ Alternating Layer Spin $\sim$ chiral density waves $\sim$ chiral spirals}

\medskip

Last word on these issues, certainly, has not been said. The modern quarkyonic interpretation of the constituent quark matter, with the 
Polyakov line and large-$N_c$ arguments offers, along the quest of exploring the 
QCD phase diagram, the ground for studying nonuniform phases,. 

%\bibliographystyle{h-elsevier3}
%\bibliography{quarky}

\end{document}